# RECOGNIZING THE VALUE OF THE SOLAR GRAVITATIONAL LENS FOR DIRECT MULTIPIXEL IMAGING AND SPECTROSCOPY OF AN EXOPLANET

National Academy of Sciences
The Committee on an Exoplanet Science Strategy Call for White Papers

**A WHITE PAPER**

Slava G. Turyshev[1], Michael Shao[1], Janice Shen[1], Hanying Zhou[1], Viktor T. Toth[2], Louis Friedman[3], Leon Alkalai[1], Nitin Arora[1], Darren D. Garber[4], Henry Helvajian[5], Thomas Heinsheimer[5], Siegfried W. Janson[5], Les Johnson[6], Jared R. Males[7], Roy Nakagawa[5], Seth Redfield[8], Nathan Strange[1], Mark R. Swain[1], David Van Buren[1], John L. West[1], and Stacy Weinstein-Weiss[1]

[1]*Jet Propulsion Laboratory, California Institute of Technology, 4800 Oak Grove Drive, Pasadena, CA 91109-0899, USA*

[2]*Ottawa, Ontario K1N 9H5, Canada, www.vttoth.com*

[3]*The Planetary Society, Emeritus*

[4]*NXTRAC Inc., Redondo Beach, CA 90277, https://nxtrac.com/*

[5]*The Aerospace Corporation, El Segundo, CA*

[6]*NASA George C. Marshall Space Flight Center, Huntsville, AL*

[7]*Department of Astronomy & Stewart Observatory, University of Arizona, Tucson AZ 8572, USA*

[8]*Wesleyan University, 96 Foss Hill Drive, Middletown, CT 06459, USA*

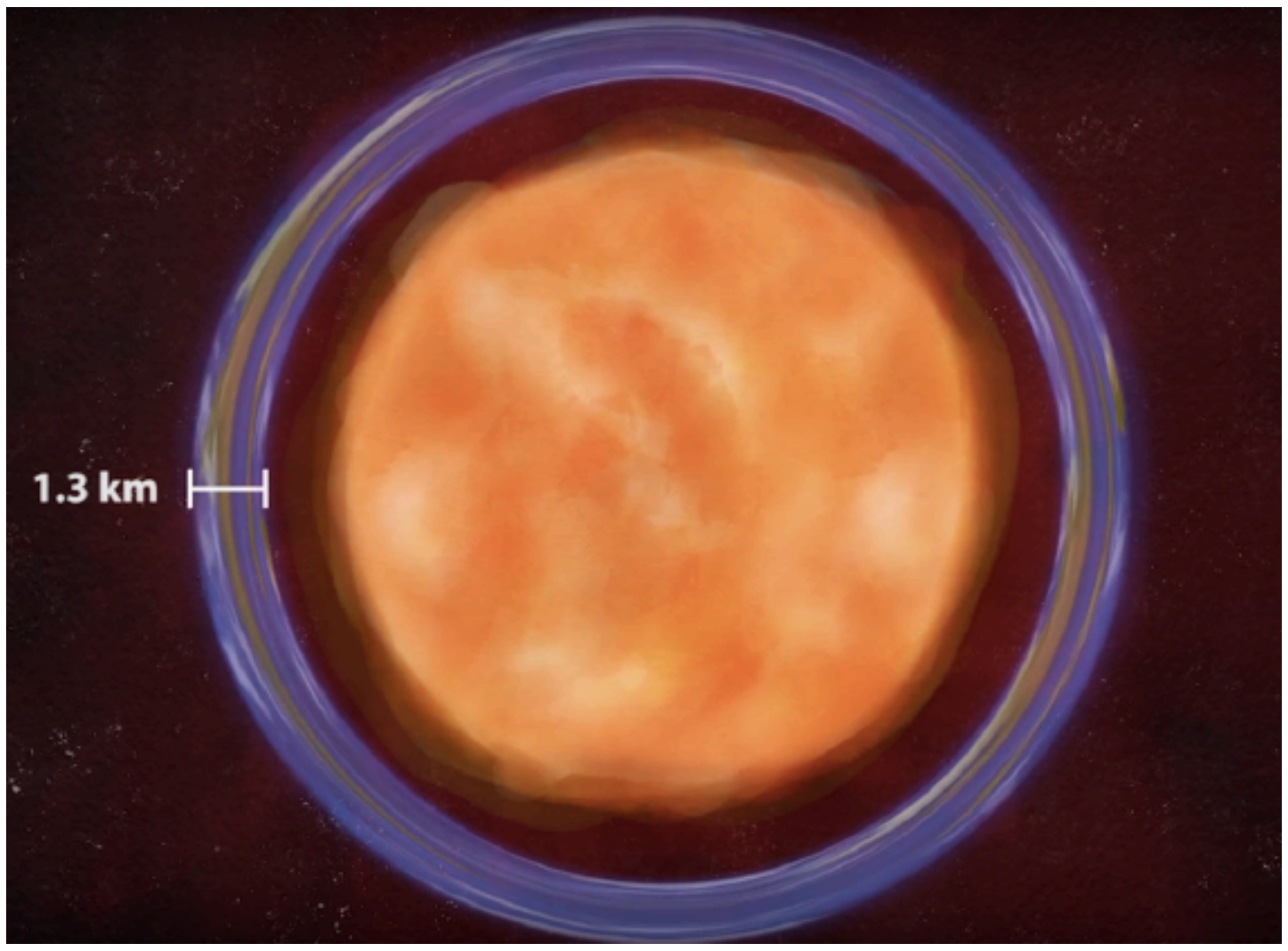

Einstein ring with thickness of 1.3 km around the Sun formed by an exo-Earth (not to scale!). A meter-class telescope, with a modest coronagraph to block solar light with $10^{-6}$ suppression placed in the focal area of the solar gravitational lens, can image an exoplanet, say at a distance of 30 parsec, with few kilometer-scale resolution on its surface. Notably, spectroscopic broadband SNR is ~$10^6$ in two weeks of integration time, providing this instrument with incredible remote sensing capabilities. See conceptual video description at https://youtu.be/Hjaj-Ig9jBs

Slava G. Turyshev, (818) 393-2600, turyshev@jpl.nasa.gov

## 1 Introduction

Direct detection of light from an Earth-like exoplanet in a habitable zone orbit is a challenging task. The angular size of this object is very tiny, requiring extremely large apertures or interferometric baselines. The light contamination from the parent star necessitates using advanced coronagraphic techniques. The light received from the exoplanet is exceedingly faint and rides on a noisy background. Detecting such a signal requires stable pointing and very long integration times. These challenges make direct high-resolution imaging of an exoplanet with a conventional telescope or interferometer a very difficult, if not impossible task.

Fortunately, Nature has presented us with a powerful instrument that we have yet to explore and learn to use for imaging purposes. This instrument is the Solar Gravitational Lens (SGL), which takes advantage of the natural ability of our Sun's gravitational field to focus and greatly amplify light from faint, distant sources of significant scientific interest, such as a habitable exoplanet.

According to Einstein's general theory of relativity, the gravitational field induces refractive properties on spacetime. A massive object acts as a lens, bending the trajectories of incident photons (Turyshev & Toth, 2017). As a result, gravitationally refracted rays of light passing on two sides of the lensing mass converge. A gravitational lens has spherical aberration, with the bending angle being inversely proportional to the impact parameter of the light ray with respect to the lensing mass. Therefore, such a lens produces not a single focal point but a semi-infinite focal line.

Although all the solar system's bodies act as lenses, only the Sun is massive and compact enough for the focus of its gravitational deflection to be within the range of a realistic mission. Its focal line is broadly defined as the area beyond ~547.8 AU, on the line connecting the center of an exoplanet and that of the Sun. A probe positioned beyond this heliocentric distance could use the SGL to magnify light from distant objects on the opposite side of the Sun (Eshleman, 1979).

While all currently envisioned NASA exoplanetary concepts would be lucky to obtain a single-pixel image of an exoplanet, a mission to the focal area of the SGL, carrying a modest telescope and coronagraph, opens up the possibility for *direct* imaging with $10^3 \times 10^3$ pixels and high-resolution spectroscopy of an Earth-like planet. Thus, an exoplanet at a distance of 30 parsecs (pc) may be imaged with a resolution of ~10 km on its surface, enough to see its surface features and signs of habitability, as shown in the video (DeLuca, 2017).

The remarkable optical properties of the SGL include major brightness amplification ($\sim 10^{11}$ at the near-infrared wavelength of $\lambda = 1$ μm) and extreme angular resolution ($\sim 10^{-10}$ arcsec) within a narrow field of view (Turyshev & Toth, 2017). A modest telescope at the SGL could be used for direct imaging of an exoplanet. The entire 13,000 km image of such an exo-Earth is projected by the SGL into an instantaneous cylindrical volume with a diameter of ~1.3 km surrounding the focal line. Moving outwards while staying within this volume, the telescope will take photometric data of the Einstein ring around the Sun, formed by the light from the exoplanet. The collected data will be processed to reconstruct the desirable high-resolution image and other relevant information.

Recently, we have evaluated the feasibility of the SGL for direct imaging and spectroscopy of an exoplanet (Turyshev et al., 2018). While several practical constraints have been identified, there appear no fundamental limitations either with concept or the technologies needed for the mission. The investigation primarily analyzed the requirements of how to operate a spacecraft at such enormous distances with the needed precision. Specifically, we studied i) how a space mission to the focal region of the SGL may be used to obtain high-resolution direct imaging and spectroscopy of an exoplanet by detecting, tracking, and investigating the Einstein ring around the Sun, and ii) how such information could be used to detect if there are signs of life on another planet.

Most importantly, we determined that the foundational technologies already exist and some of them have a rather high technology readiness level. With the advent of new launch capabilities (Space Launch System, Falcon 9 Heavy, etc.) and recent advances in spacecraft technologies (electric propulsion, solar sails, optical communications, etc.), there is clear motivation to further explore the utility of the SGL for remote investigation of exoplanets.

## 2 Technical Description

***Theory:*** A wave-theoretical description of the SGL (Turyshev & Toth, 2017) demonstrates that it possesses a set of rather remarkable optical properties. Specifically, the SGL amplifies their brightness of light from distant, faint sources by a factor of ~$2GM/(c^2\lambda)$ ~ $10^{11}$ (for $\lambda = 1$ μm), which is an enormous magnifying power not easily feasible with conventional astronomical instruments. Moreover, the SGL has extreme angular resolution of $\lambda/D_0$ ~$10^{-10}$ arcseconds (with $D_0$ being the diameter of the Sun), which makes it exceptionally well-suited for imaging distant objects.

The angular diameter of an Earth-like exoplanet at 30 pc is $1.4 \times 10^{-11}$ rad. To resolve the disk of this planet as a single pixel, a telescope array with a baseline of ~74.6 km would be needed. Resolving the planet with $10^3$ linear pixels would require a baseline ~$1\times10^5$ km (~$12R_\oplus$), which is not feasible.

In contrast, a 1-m telescope, placed at the focal line of the SGL at 750 AU from the Sun, has a collecting area equivalent to a diffraction-limited telescope with diameter of ~80 km and the angular resolution of an optical interferometer with a baseline of $12R_\oplus$. While building an instument with the SGL's capabilities is far beyond our technological reach, we can use the SGL's unique capabilities to capture megapixel images of exoplanets.

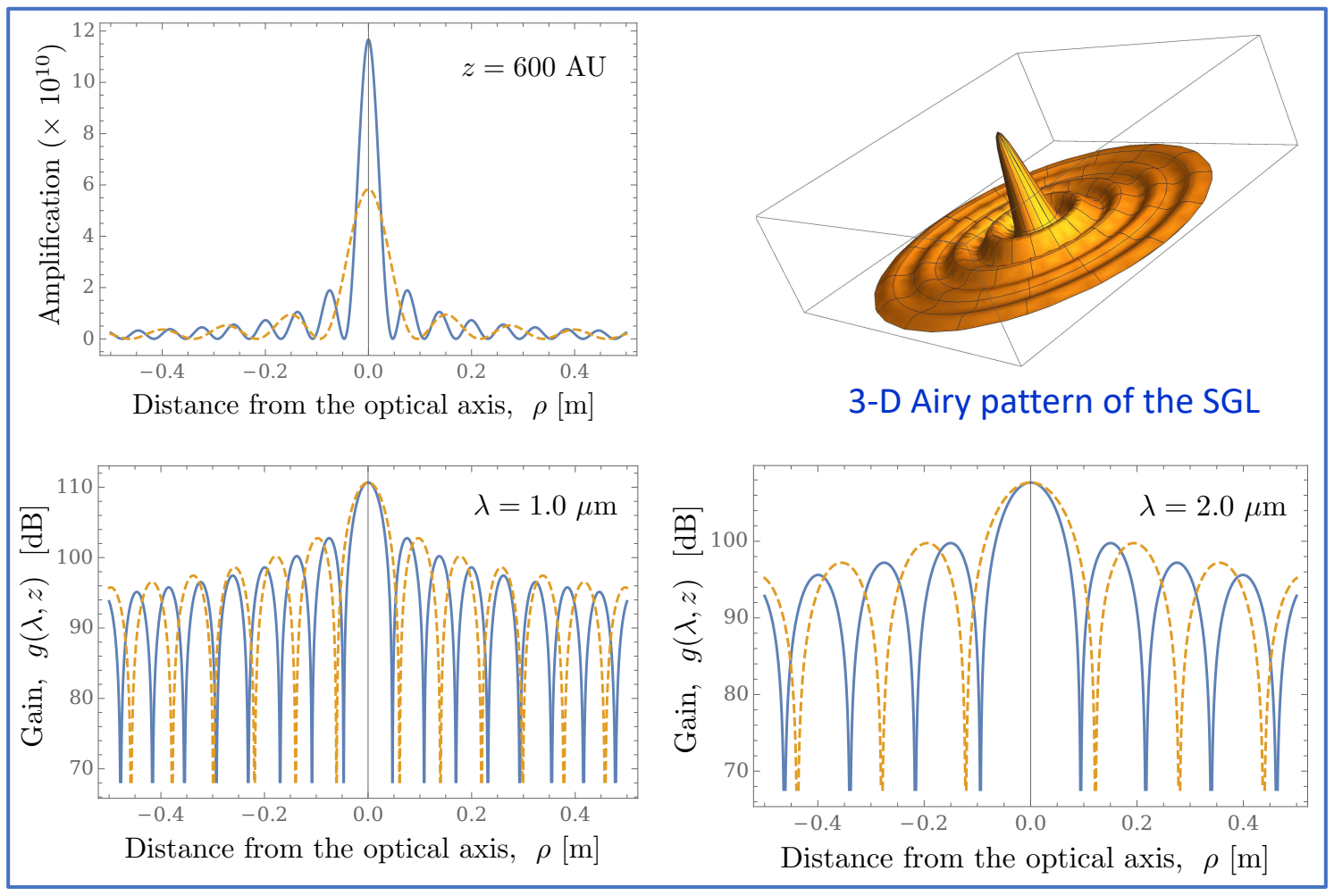

Figure 1. Optical properties of the SGL (Turyshev & Toth, 2017). Up-Left: Amplification of the SGL. Up-Right: Point spread function. Bottom: Gain of the SGL as seen in the image plane as a function of possible observational wavelength.

Recent efforts have produced a better understanding of the optical properties of the SGL. Figure 1 shows the lens' point-spread function (PSF), resolution, magnification, all of which helps in the design of the large scale astronomical facility that would benefit from the SGL.

***Imaging concept:*** The image of an exo-Earth, say at 30 pc, is compressed by the SGL to a cylinder with a diameter of ~1.3 km (corresponding to the Einstein ring around the Sun with the same 1.3 km thickness) in the immediate vicinity of the fictitious focal line. Imaging an exo-Earth with $10^3 \times 10^3$ pixels requires moving the spacecraft within the image plane in steps of 1.3 km $/10^3$ ~ 1.3 m while staying within the ~1.3 km diameter cylindrical volume. So, each ~1 m pixel in the image plane corresponds to a pixel diameter of $13\times10^3$ km$/10^3$ ~ 10 km on the surface of the planet.

The challenge comes from recognizing the fact that the PSF of the SGL is quite broad (see Figure 1), falling off much slower than the PSF of a typical lens. Consequently, for any pixel in the image plane, this leads to combining light from a particular pixel on the surface of the exoplanet but also from many pixels adjacent to it. This admixing of light leads to a significant blurring of the image.

To overcome this impediment, imaging must be done on a pixel-by-pixel basis by measuring the brightness of the Einstein ring at each of the image pixels. The knowledge of the PSF's properties, makes it possible to apply deconvolution algorithms that enable reconstruction of the original image efficiently. To make this process work, a significant signal to noise ratio (SNR) is required. Fortunately, the SGL light amplification capability provides a SNR of over $10^3$ given a 1 second of integration time. This is sufficient for nearly noiseless deconvolution.

Light contamination from the parent star is a major problem for all modern planet-hunting concepts. However, for the SGL, due to its ultrahigh angular resolution ($\sim 10^{-10}$ arcsec) and very narrow field of view, the parent star is completely resolved from the planet with its light amplified $\sim 10^4$ km away from the planet's optical axis, making the parent star contamination issue negligible.

***Instrument:*** Thanks to the large photometric gain of the SGL, its high angular resolution and strong spectroscopic SNR ($10^3$ in 1 sec), a small diffraction-limited high-resolution spectrograph is sufficient for the unambiguous detection of life (Turyshev et al., 2018).

As the instrument ultimately determines the size of the spacecraft and, thus, its motion in the image plane, recently we addressed the issues of coronagraph design. For this, we require the coronagraph to block solar light to the level of the solar corona brightness at the location of the Einstein ring.

At 1 μm, the light amplification of the SGL is $\sim 2 \times 10^{11}$ (equivalent to −28.2 mag), so an exoplanet, which initially is seen as an object of 32.4 mag now becomes a ~4.2 mag object. However, when averaged over a 1-m telescope, light amplification is reduced to $\sim 2\times10^9$ (−23.25 mag), the exoplanet becomes a 9.2 mag object, still sufficiently bright. However, the image will include noise in the form of light from the solar corona, the residual solar light, and the zodiacal light.

To validate our design assumptions, we performed a preliminary coronagraph design and simulations. Suppressing the Sun's light by a factor of $10^{-6}$ when imaging with the SGL is significantly less demanding than the requirements for modern-day exoplanet coronagraphs, which must suppress the parent star's light by a factor of $10^{-10}$ to detect an exo-Earth at least as a single pixel.

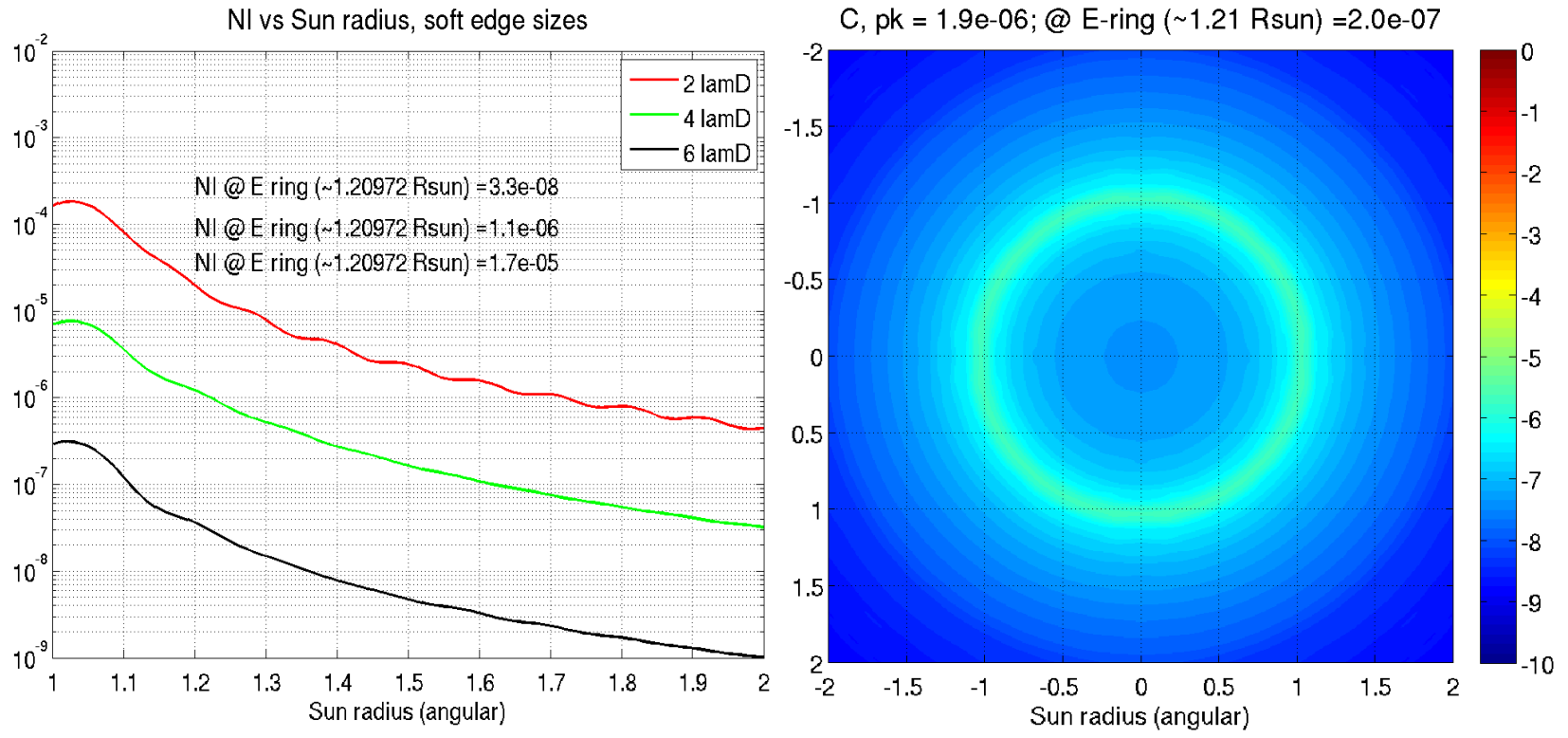

Figure 2. Left: Gaussian soft-edge has a great impact on light suppression ability of the coronagraph. Right: Simulated coronagraph performance showing the solar light suppression by $2\times10^{-7}$, sufficient for imaging with the SGL.

We evaluated the performance of the coronagraph with a Fourier-based diffraction model. The Sun is modelled as a dense collection of incoherent point sources with its corona at the relevant heliocentric ranges obeying $\sim r^{-3}$ power law profile. Design parameters include telescope size, distance to the SGL, occulter mask profile and Lyot mask size. The full width at half maximum of the Gaussian soft edge has a significant impact on the coronagraph's performance (Figure 2).

Defining contrast as brightness normalized to peak brightness without coronagraph, we achieved a total planet throughput of ~10%. Figure 2 shows the contrast at the image plane after the coronagraph. At a contrast of $2 \times 10^{-7}$, the leaked solar light is ~5 times lower in intensity than the corona, satisfying the stated objectives for imaging with the SGL (Turyshev et al., 2018).

***Image Reconstruction:*** Creating a megapixel image requires $\sim 10^6$ separate measurements. For a typical CCD photography, each detector pixel within the camera is performing a separate measurement. This is not the case for the SGL. Only the pixels in the telescope detector that image the Einstein ring measure the exoplanet, and the ring contains information from the entire exoplanet, due to the disproportional image bluring by the SGL and also due to the relative distribution of different regions of the exoplanet to different azimuths of the ring.

To evaluate the imaging performance of the SGL, we used a technique of rotational deconvolution. The idea is that the Einstein ring seen by the imaging telescope is actually a collection of ringlets formed by light from individual pixels on the exoplanetary surface. Each of these ringlets contains a highly aberrated image of that particular surface pixel. Thus, the entire Einstein ring contains highly blurred light from the entire planet. The planetary rotation allows for different parts of the planet to rotate in/out of the view, leading to brightness variations. Thus, one can reconstruct an albedo map of the planet by using 1-day light curves while performing longitudinal deconvolution.

To implement this approach, we used images of the Earth taken by the EPIC camera on the DSCO[1] spacecraft at the Sun-Earth L1 Lagrange point. The raw data is available as near real-time images of Earth[2]. The input data was used to produce an albedo map parameterized by longitudes and latitudes. The half of the planet that is visible at any given time is integrated to produce a light curve representing the variation of the total flux, as would be seen by the SGL (Turyshev et al., 2018).

This light curve was inverted using a straightforward pseudo-inverse procedure, producing a longitudinal map of the planet. If the planet's spin axis is not aligned with its orbital axis, then over the course of a year the illumination over latitude also changes. Figure 3 shows the input albedo map and the output of the 2D rotational deconvolution. This type of deconvolution requires a very high photometric SNR. The SNR loss during deconvolution is roughly proportional to the number of pixels. For instance, given 100 pixels in latitude and 300 in longitude, the resulting $3 \times 10^4$ pixel albedo map will have $\sim 3 \times 10^4$ hits on the photometric SNR. Fortunately, the very large effective collecting area of the SGL makes possible a sufficiently large SNR.

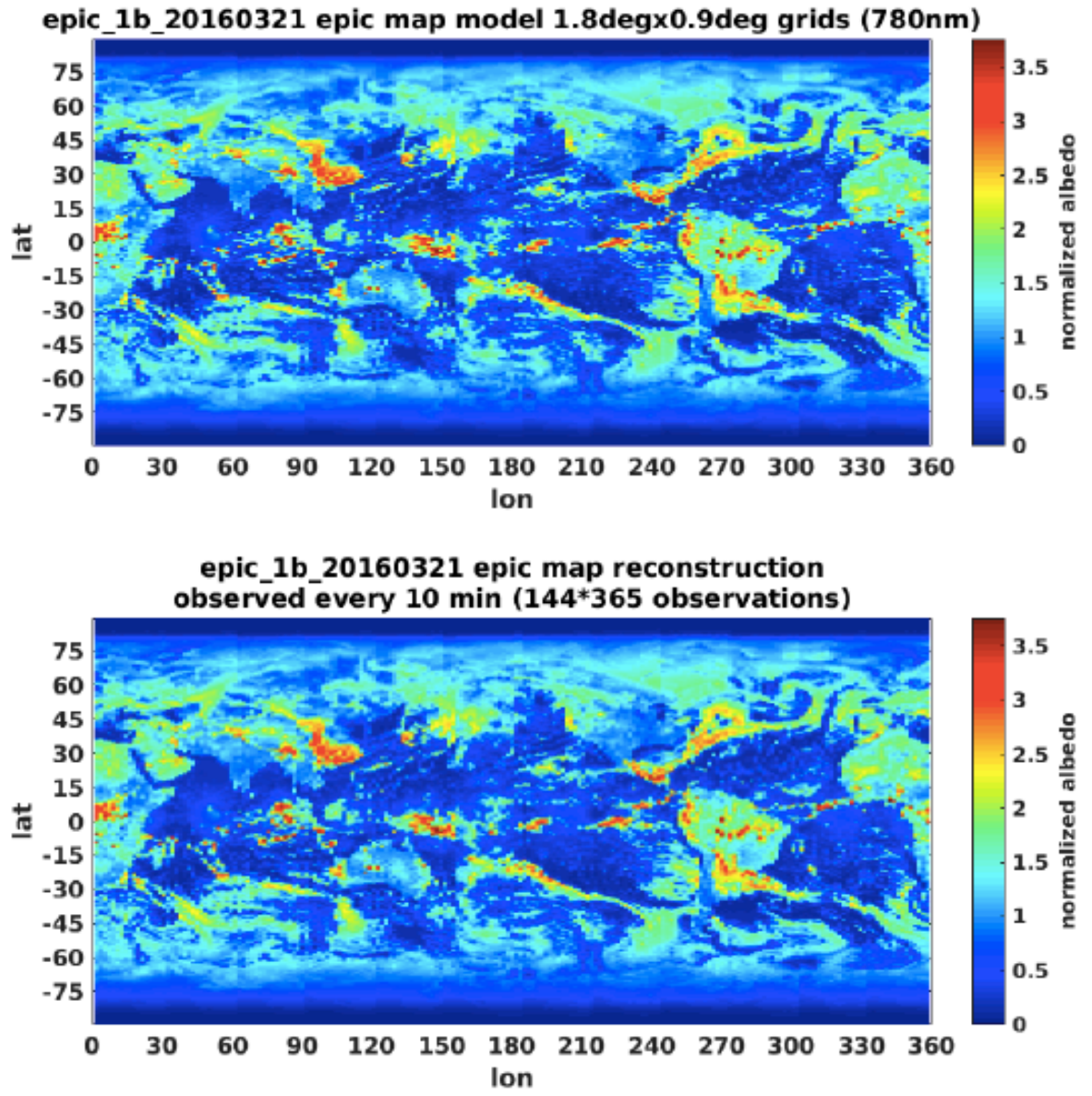

Figure 3: Rotational deconvolution: Top is the input map, which is the albedo map of the Earth used to generate light curves. The bottom is the albedo map calculated by "inverting" the light curves back to a 2D image.

What is encouraging is that the temporal variability in cloud cover helps the deconvolution. Assuming $N \sim 50$ observations of every pixel, the effect of clouds is suppressed after ~10 observations. If spectroscopic data is also used, the contribution of cloud cover can be further suppressed, allowing us to "see through" the clouds. The rotational deconvolution of the EPIC camera data lets us see the surface of the Earth using data comprising a few months.

With direct deconvolution, we show that with a 1-m telescope, it would take ~2 years to build a $500 \times 500$-pixel image. Two factors that can reduce the integration time by a factor of up to 100 are i) the number of image pixels, $N$, and ii) the telescope diameter. The higher the desirable resolution the longer is the integration time, $T$, scaling as $T \sim N D^4$, where $D$ is the distance to the target. Another scaling law is related to the telescope diameter, $d$. A telescope with double the aperture will collect four times as many photons. Its diffraction pattern will be twice as narrow and, thus, it will collect half as many photons from the solar corona photons. The integration time scales as $T \sim d^{-3}$. Thus, a larger image of $10^3 \times 10^3$ pixels may be produced in ~4 years if a 2-m telescope is used. However, $T$ may be reduced if there are time-varying features in the planetary albedo (regular features and/or cloud pattern, etc.). The time is also reduced by $\sim n^{-1}$ if $n$ imaging spacecraft are used. Other helpful factors are i) the rotational motion of a crescent exo-Earth, and ii) increase in heliocentric distance and the resulting improvements in coronagraphic performance.

[1] Deep Space Climate Observatory (or DSCOVR), https://en.wikipedia.org/wiki/Deep_Space_Climate_Observatory

[2] Website for Earth Polychromatic Imaging Camera (EPIC), https://epic.gsfc.nasa.gov/

We have shown (Turyshev et al, 2018) that imaging with the SGL, although complex, has no fundamental "show-stoppers." Given the enormous light amplification provided by the SGL, spectroscopic investigations, even spectro-polarimetry could be viable. Ultimately, what could be obtained is not just an image, but a spectrally-resolved image over a broad range of wavelengths, characterizing the atmosphere, surface materials and biological processes on that exo-Earth.

## 3 Conclusion

By detecting numerous potential Earth-like exoplanets, the Kepler mission has raised the possibility of existence of another Earth-like world. Follow-ups on Kepler candidates with other current exoplanet characterization technologies may yield unresolved images at low spectral resolution (typically $R<100$). The next steps include TESS (2018), which will extend Kepler's work by performing an all-sky survey to find additional exoplanets, including Earth-like ones; JWST (2019), which will be used for targeted follow-up on candidate planets. There are also missions yet in formative stages, such as the Exo-C, Exo-S, HabEx, and LUVOIR concepts (Turyshev et al, 2018).

However, there is no concept for direct multi-pixel imaging of an exoplanet. All the exoplanet imaging concepts currently studied by NASA, aim to capture light from an unresolved Earth-like exoplanet as a single pixel. These missions would provide globally averaged measurements of the atmosphere, identify major biomarkers, etc. But, SGL will open up scientific questions to the exoplanet community that are currently only possible by planetary scientists with the solar system planets (e.g., studying surface landforms to evaluate the geologic evolution of the planet). In addition, a spatially resolved spectroscopic image allows us to probe small structures and detect weak features that would be lost in a global average (e.g., surface volcanism, land/water interactions, spatially limited biosignatures). Also, the SGL provides the opportunity to make a direct detection of life, as opposed to the indirect detection from a globally averaged spectroscopic biomarkers.

This work is of major interdisciplinary importance. Just getting out to the SGL provides lots of interesting collaborations with heliophysics and astrophysics (Stone et al., 2015). Also, there is more direct interdisciplinary nature of planetary scientists and exoplanet scientists. Although, this connection is growing there is still a major chasm between the rich observations that planetary scientists use versus the few globally averaged parameters that astronomers are using with exoplanets.

We note that mission to the SGL is a long term objective that requires some planning and work now with some of the technical issues involving propulsion, communication, autonomy (Turyshev et al., 2018). However, once/if we find Earth-like planets with biosignatures, a spatially resolved spectroscopic observation is THE imperative next step and the easiest path to that is the SGL.

Therefore, we ask the NAS Committee on an Exoplanet Science Strategy to consider mission concepts capable of exploiting the remarkable optical properties of the SGL for *direct* high-resolution imaging/spectroscopy of a potentially habitable exoplanet. Such missions could allow exploration of exoplanets relying on the SGL capabilities decades, if not centuries, earlier than possible with other extant technologies. We also ask that the Committee recommends NASA to consider investing in technologies needed for missions to reach and operate at the relevant heliocentric distances.